\documentclass[aps,twocolumn,prb,reprint,amsmath,amssymb,showpacs,superscriptaddress]{revtex4-1}

\usepackage{graphicx}
\usepackage{dcolumn}
\usepackage{bm}
\usepackage{amsmath}
\usepackage{color}
\usepackage{mathrsfs}
\usepackage{float}
\usepackage{dsfont}
\usepackage{txfonts}
\usepackage{wasysym}
\usepackage{ifsym}

\makeatletter

\newcommand{\Rmnum}[1]{\expandafter\@slowromancap\romannumeral #1@}
\makeatother
\begin{document}

\title{Anomalous Joule law in the  adiabatic dynamics of a normal-superconductor quantum dot} 

\author{Liliana Arrachea}
\affiliation{International Center for Advanced Studies, Escuela de Ciencia y Tecnolog\'{\i}a, Universidad Nacional de San Mart\'{\i}n-UNSAM, Av 25 de Mayo y Francia, 1650 Buenos Aires, Argentina} 
\affiliation{Dahlem Center for Complex Quantum Systems and Fachbereich Physik, Freie Universit\"at Berlin, 14195 Berlin, Germany}
\author{Rosa L\'opez}
\affiliation{x}


\begin{abstract}
We formulate a general theory to study the time-dependent charge and energy transport of 
an adiabatically driven quantum dot in contact to  normal and superconducting reservoirs at $T=0$. This setup is a generalization of a quantum RC circuit, with capacitive components due to Andreev processes and induced pairing fluctuations, in addition to the convencional  normal charge fluctuations. The dynamics for the dissipation of energy is ruled by a Joule law of four channels in parallel with the
universal B\"uttiker resistance $R_0=e^2/2h$ per channel. Two transport channels are
 associated to the two spin components of the usual charge fluctuations, while the other two are associated to electrons and holes due to pairing fluctuations. The latter leads to an
 "anomalous" component of the Joule law and take place with a  vanishing net current due to the opposite flows of electrons and holes.
\end{abstract}

\date{\today}

\maketitle


 \section{Introduction}
Time dependent transport at nanoscale is a prominent tool for probing electronic dynamics at very low temperatures.
 A prototypical instance is found in on-demand single electron sources in which individual electron 
and hole charges are perfectly emitted. \cite{qcap2} The simplest device that works as a quantized emitter is a quantum capacitor, which
consists of a single-level quantum dot tunnel-coupled to an unique reservoir. In such a case only a purely AC current response is 
possible when the dot gate is electrostatically influenced by an AC voltage source. \cite{btp1,btp2,btp3,qcap1,qcap3,qcap4} Working in a range 
of frequencies of GHz ($\Omega$) and at sufficiently slow AC amplitudes ($V_g$) this setup behaves as a RC circuit that for the quantum regime exhibits the 
peculiarity that relaxation processes are featured by an universal quantized resistance $R_0=h/2e^2$. \cite{btp1,btp2,btp3} The quantum analogue to the classical RC circuit is now done by replacing the geometrical capacitance by  a quantum capacitance which is proportional to the density of states of the localized level.  

Conductance quantization is observed in the stationary regime as a signature of ballistic transport due to the 
lack of backscattering events. 
\cite{Wees88,Klit80}  In a quantum capacitor operating in conditions where many-body interactions do not play a role, the resistance quantization is attributed to a particular behavior 
of the dwell time. $R_0$ is universal because the mean value for the square of the dwell time coincides with the square of its mean 
value.   For interacting systems under AC driving  charge relaxation processes are dictated by the correlation function of the electron-hole excitations 
which are proportional to the available density of electron-hole pairs or, equivalently,  to the  charge susceptibility.  \cite{mora, rosa1, ringel, hama, rosa2,
mich1,rossello,dut,janine,mich1,mich2,mich3} In that case, there is a relaxation resistance $R_0$ per
spin channel and 
such 
universality resides in the fullfilment of the Korringa-Shiba relation. \cite{korr-shiba,mich1,mich2,mich3} The latter holds for 
systems that behave as Fermi liquids, which  to some extend behave as noninteracting systems with renormalized parameters. 
Besides, a different quantization phenomenon in a quantum capacitor is observed, depending on the way in which the AC amplitude is increased beyond linear response. 
\cite{nonlin,qcap1,qcap3,qcap4, misha1,misha2,misha3,misha4,keel,piet}
 Such quantization has potential metrological 
applications and is suitable for quantum computing designs. 
Most of the studies on quantum RC circuits belong to the linear regime being the nonlinear regime less investigated.  In particular, few studies have been reported in the interacting system beyond linear response. \cite{alei,piet,david,rome}

 \begin{figure}
\includegraphics[width=\columnwidth]{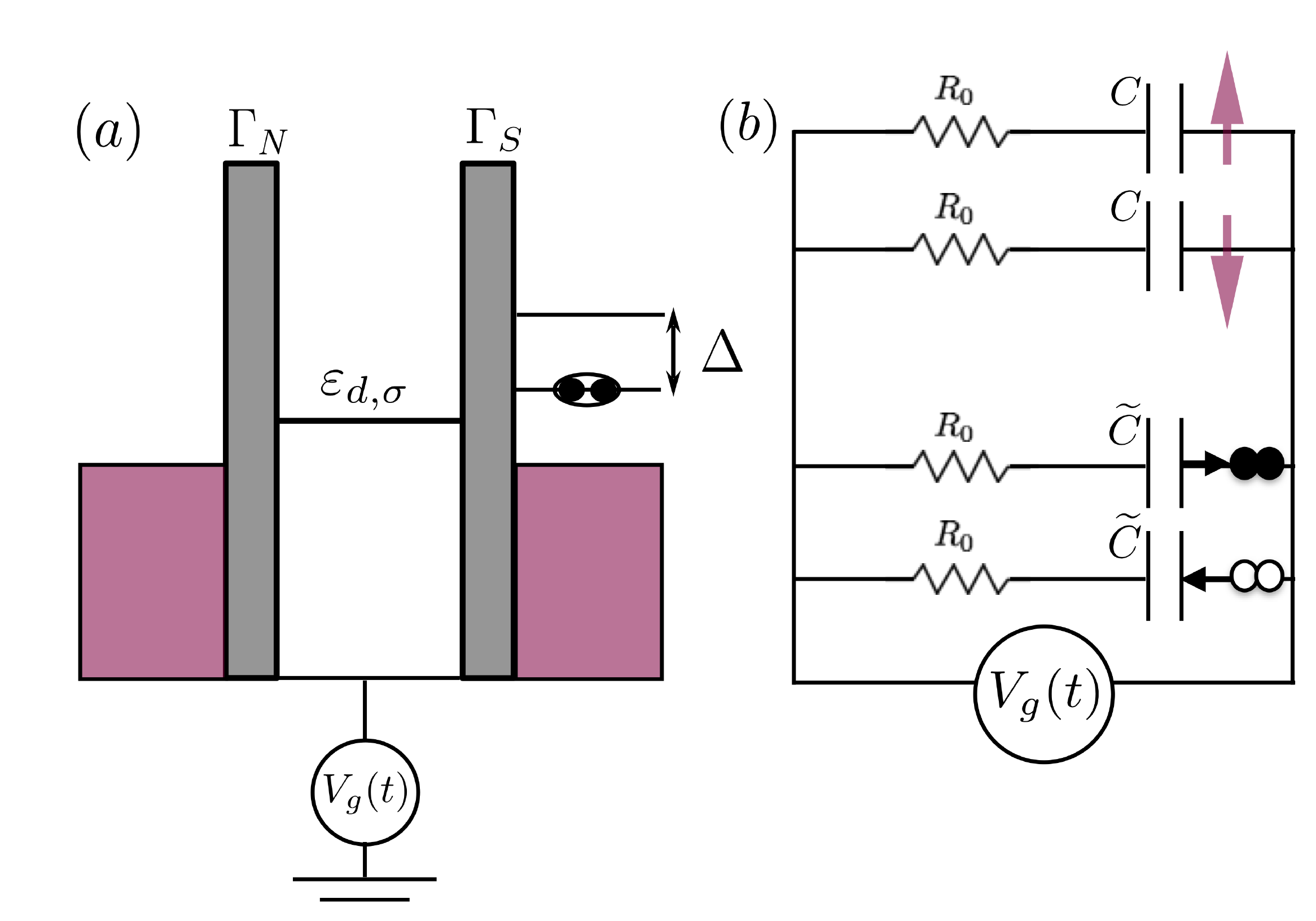}
\caption{ (a) Sketch of the setup.  A quantum dot  is driven by an ac gate voltage $V_g(t)= V_0 \sin (\Omega t)$
and is connected to a normal and a superconducting lead. (b) Representation of the dynamics of the charge and energy by means of an effective circuit with capacitive and resistive elements. The two branches associated to the normal charge fluctuations are associated to the two components of the spin of the electrons. The other two describe the "anomalous" 
charge flow due to fluctuations of the induced pairing at the quantum dot. The corresponding charge fluxes are associated to electrons and holes and have opposite directions as indicated in the figure by the full dot pairs (electrons) and empty ones (holes).}
 \label{fig1}
\end{figure}

In the nonlinear regime, it is not obvious how to extend the concept of relaxation resistance, because 
the analogy to the classical circuit is not necessarily valid. Resistive behavior is related to dissipation of energy. Hence, the analysis of the energy transport and heat production in parallel with the charge transport in these systems is a natural strategy. 
In a recent work, it was shown that a non-interacting quantum dot driven in the adiabatic regime obeys an {\em instantaneous Joule law} with an universal resistance $R_0$ per transport channel. \cite{re1,re2,re3,re4} For an interacting quantum dot described by the Anderson impurity model, the fact that the instantaneous susceptibility satisfies Korringa-Shiba law, ensures the validity of the same universal instantaneous Joule law. 
When a magnetic field is included in  this model for  the interacting quantum dot, the Joule law is not satisfied separately for each spin channel but it is satisfied by the effective resistance of the two spin channels
considered in a parallel circuit configuration. \cite{rome}

A very interesting generalization of the RC circuit is to consider a configuration where the capacitive element --the quantum dot-- is not only connected to a conducting lead but also to a superconducting one. DC transport in setups containing a quantum dot embedded in a normal-superconductor (N-S) junction  has been widely investigated theoretically and also experimentally. \cite{qdot1,qdot2,qdot3,qdot4,qdot5} Pumping induced by AC driving  in quantum dots in N-S junctions has also been investigated. \cite{qdotp1,qdotp2,qdotp3} However, only  recently RC configurations and time-dependent transport induced by a single driving potential at the quantum dot in these structures have been studied. \cite{qdots} The extra ingredient that the N-S 
coupling brings about is the conversion of electron-hole pairs into Cooper pairs between the two leads because of the Andreev processes, \cite{Andreev} along with induced superconductivity at the quantum dot. The aim of the present work is to explore the impact that these effects have in the interplay between the charge and the energy dynamics of such  hybrid RC setups.
A sketch of the setup is shown in Fig. \ref{fig1}. We will focus on the adiabatic regime, where the period of the driving gate voltage is much larger than any characteristic time for the electrons in
the quantum dot and both leads  are at temperature $T=0$.
We will show that Andreev processes introduce an additional contribution to the quantum capacitance, $C_{\rm And}(t)$, induced by the coupling to the normal lead, $C_{\rm N}(t)$, while the induced pairing due to the coupling to the superconducting lead can be represented by an anomalous capacitance $\tilde{C}(t)$. The latter  describes the simultaneous fluctuations
of electrons and holes associated to the fluctuation of the induced pairing, as a response to a variation of the gate voltage. Each of these capacitances depend on time in the regime where
the amplitude of the driving voltage exceeds the range of linear response. The concomitant energy conversion can be described by an instantaneous Joule law. The latter is a generalization
of the Joule law of Refs. \onlinecite{re1,re2,re3}, where, in addition to the contribution of the two spin channels, there is an anomalous component due to the disruption and formation of the
induced pairing at the quantum dot. Unlike the former contribution, the latter takes place without a net charge flow, since electrons and holes generate currents in opposite directions. The corresponding  processes can be represented by the circuit of Fig. \ref{fig1}. The paper is organized as follows. In section II we present the model. Section III contains the equations ruling the charge and energy dynamics, including the introduction of the adiabatic regime and the Green's function treatment to calculate the relevant time-dependent observables. The instantaneous Joule law is derived from the quantum-dot dynamics in Section IV, while in Section V we show that the associated heat flows entirely into the normal lead. In Section VI we present some results
that illustrate the behavior of the different components of the capacitances and the different components of the Joule heating. Finally, summary and conclusions are presented in Section VII.

\section{Model}
We consider a single-level quantum dot that is tunnel-coupled to both, a superconducting (S) and normal (N) reservoirs.  The quantum dot is under the action of an oscillatory time-dependent gate potential 
$V_g(t)= V_0 \sin (\Omega t)$. The full set-up is described by the Hamiltonian, 
\begin{equation}
\label{Hamtot}
H(t)= H_{\textrm{d}}(t)+H_{\textrm{N}} + H_{\textrm{S}}  + \sum_{\alpha= {\rm N, S}} H_{\textrm{c} \alpha }.
\end{equation}
The first term describes a single level quantum dot
\begin{equation} \label{hd}
H_{\textrm{d}} (t) =  \sum_{\sigma} \left[\varepsilon_{d, \sigma}+ e V_g(t)\right]n_{d \sigma},
\end{equation}
where $d^\dagger_\sigma$ is the creation operator for an electron on the dot with spin $\sigma=\uparrow,\downarrow$, and  $n_{d\sigma}$ denotes the occupation operator for spin $\sigma$. 
$\varepsilon_{d,\sigma}$ is the energy of the dot  level, which is modulated by $V_g(t)$.
 The normal reservoir is described by a free-electron Hamiltonian 
 \begin{equation}\label{hn}
H_{\rm N} =\sum_{\sigma,k{\rm N}}(\epsilon_{k{\rm N}}-\mu_N)c_{k{\rm N},\sigma}^{\dagger}c_{k{\rm N},\sigma},
\end{equation}
 in which $\epsilon_{k{\rm N}}$ is the energy dispersion relation and $k$ the wavevector, $c_{k{\rm N},\sigma}$ is the destruction operator for 
 an electron in the normal reservoir with spin $\sigma$. The electrochemical potential for the normal contact is 
 represented by $\mu_N$.  The superconducting reservoir is described by a BCS Hamiltonian of the form 
\begin{equation}\label{hs}
H_{\rm S} =\sum_{\sigma,k{\rm S}}\left[(\epsilon_{k {\rm S}} -\mu_S)c_{k {\rm S},\sigma}^{\dagger}c_{k {\rm S},\sigma}
+ \Delta c_{k {\rm S} \uparrow} c_{-k {\rm S}, \downarrow} + h.c.\right]\,,
\end{equation}
where $\Delta$ denotes the s-wave pairing potential.  The coupling between dot and reservoirs is 
\begin{equation}
H_{\textrm{c} \alpha}= w_{\alpha}\sum_{k\alpha, \sigma}\left[c_{k\alpha, \sigma}^{\dagger}d_{\sigma} +h.c\right]\,,
\end{equation}
Here, $w_{\alpha}$ is the tunneling amplitude that connects both, the normal reservoir with the dot and the superconducting contact with the central site. We focus on the transport induced purely by the AC driving applied at the quantum dot, without any additional voltage bias applied at the leads. For simplicity, we consider $\mu_S=\mu_N=0$.

\section{Charge dynamics and dissipation}
In this section we formulate  the equations describing the charge and energy  dynamics of the full system. In the forthcoming sections we will analyze  the problem from two complementary perspectives, (i) we calculate the dot charge dynamics and the dissipated power in the adiabatic regime and we will show that both quantities are related by means of an instantaneous Joule law with a constant and universal resistance.  Such relation follows from a circuit description in which quasiparticles and pair generation events run in parallel (see Fig. 1). (ii) Secondly, we will focus on the case where the chemical potential lies within the
gap of the S reservoir.  Under these conditions, we calculate the heat flow at the normal contact and the charge current flow at the same lead. Again, we show a relationship between these two quantities
given by an instantaneous Joule law.
Remarkably, we arrive at this conclusion by evaluating the heat flow at the normal contact considering the contribution from the tunneling barrier, the energy reactance. \cite{re1,re4}

\subsection{Charge and energy dynamics of the quantum dot}
The quantum dot charge dynamics determines not only the charge current but also the amount of dissipated energy in the hybrid setup. Such dynamics is governed by a conservation law  for the electrical charges. In this respect, the flow of charges across the quantum dot fullfils 
\begin{equation}\label{ic}
e \dot{n}_{d}(t)=e \sum_{\sigma} \dot{n}_{d \sigma}(t) = - \left[ I_{\rm N}(t)+I_{\rm S}(t) \right],
\end{equation}
where $\dot{n}_{d \sigma}(t) \equiv -i /\hbar \langle [n_{d \sigma}, H] \rangle$ is the change in the occupation of the dot at time $t$ corresponding to the spin $\sigma$ and $e>0$ the electron charge. The charge currents flowing into the normal (N) and
superconducting (S) leads are computed from the Heisenberg relation, they are respectively,
\begin{equation}\label{ial}
I_{\alpha}(t)= -ie /\hbar \langle [N_{\alpha}, H] \rangle,
\end{equation}
 with
$\alpha={\rm N}, {\rm S}$ and $N_\alpha=\sum_{k\alpha,\sigma}c_{k\alpha\sigma}^\dagger c_{k\alpha\sigma}$ being the  occupation operator for the normal and superconducting contacts. 

The power supplied  by  the ac source is converted in electrical work done by the electrons at a rate
\begin{equation}\label{pac}
P(t) = - \Biggr\langle \frac{\partial H}{\partial t} \Biggr\rangle=  -  e \sum_{\sigma} n_{d \sigma}(t) \dot{V}_g(t).
\end{equation}
This power equals the total heat production rate at time $t$, \cite{re2} 
\begin{equation} \label{qtot1}
\dot{Q}_{\rm tot}(t)= - P(t)\,.
\end{equation}

\subsection{Spacial distribution  of the  heat flow}
 As explained in Refs. [\onlinecite{rossello}], [\onlinecite{re2}],   [\onlinecite{re3}],  
the heat flow is instantaneously distributed in the different parts of the device, i.e., at the contacts, central site and tunnel junctions,
as follows
\begin{equation}\label{qtot}
\dot{Q}_{\rm tot}(t)= \sum_{\alpha} \left[J^E_{\alpha}(t)  + J^E_{\textrm{c}\alpha }(t) \right]
+\dot{E}_{d}(t),
\end{equation}
where 
\begin{equation}
J^E_{\alpha}(t)= - \frac{i }{\hbar} \langle [H_{\alpha}, H] \rangle\,,\;\;\;\;J^E_{\textrm{c}\alpha}(t)= - \frac{i }{\hbar} \langle [H_{\textrm{c}\alpha}, H] \rangle
\end{equation}
is the energy rate change at the reservoirs $\alpha\in N,S$, and the corresponding contacts. The change of the energy 
at the central site is
\begin{equation}
 \dot{E}_{d}(t)=- \frac{i }{\hbar} \langle [H_d, H] \rangle + \Biggr\langle\frac{ \partial H}{\partial t} \Biggr\rangle.
 \end{equation}
 In Ref. \onlinecite{re1}, it was shown that, for a quantum dot connected to a normal lead, the most meaningful definition of the heat flux into the lead $\alpha$ is the one including
the  so-called "energy reactance", $J^E_{c\alpha}(t)/2$ that is half of the energy rate change at the tunneling barriers. The latter represents the energy that is temporarily stored or emitted at the tunneling barrier. We adopt that definition and write  the heat flux into the lead $\alpha$ as follows
\begin{equation} \label{qal}
 \dot{Q}_{\alpha}(t)=J^E_{\alpha}(t) +\frac{J^E_{\textrm{c}\alpha }(t)}{2}.
 \end{equation}
In the case of a dot connected to a single normal lead,  the reactance  ensures the validity of the second law of thermodynamics in the adiabatic regime, \cite{re1,re2, re3} it gives a proper description of the AC heat current spectrum in the linear response regime \cite{rossello}, and also of the transient dynamics. \cite{riku} Similarly, we can define the heat flow  into the quantum dot  \cite{re2,re3} as
\begin{equation} \label{qal}
 \dot{Q}_{d}(t)= \dot{E}_{d}(t)+ \sum_{\alpha} \frac{J^E_{\textrm{c}\alpha }(t)}{2}.
 \end{equation}
Notice that, by substituting these definitions in Eq. (\ref{qtot}) and using $  \sum_{\alpha} \left( [H_{\alpha}, H] +  [H_{\textrm{c}\alpha}, H]  \right)+ [H_d, H]  =0$, we get 
\begin{equation}
\dot{Q}_{\rm tot}(t)=\sum_{\alpha} \dot{Q}_{\alpha}(t)+ \dot{Q}_{d}(t)=-P(t), 
\end{equation}
which is, precisely, Eq. (\ref{qtot1}).

\subsection{Adiabatic dynamics}
We now focus on the so-called \emph{adiabatic regime}, where the AC time is much longer than any other associated time scale for the setup. In this respect, the electron tunneling processes occurs many times in a AC time period.  For the description of the quantum dot dynamics in this regime we follow Refs. \onlinecite{rome,adia}, where the quantum dot occupation is split in two contributions up to linear order in $\dot{V}_g(t)$. 
 The adiabatic evolution of the occupancy of the quantum dot is given by
\begin{equation}\label{nd}
n_{d \sigma}(t)= n^f_{d \sigma}(t)+ e \Lambda_{\sigma}(t) \dot{V}_g(t),
\end{equation}
where $n^f_{d, \sigma}(t) \equiv \langle n_{d \sigma} \rangle_ t$ is the snapshot
occupancy of the dot, evaluated with the exact {\em equilibrium} density matrix $\rho_t$ corresponding to the  Hamiltonian $H(t)$ frozen at the time $t$. The correction  is linear in both the  time variation of the AC amplitude and, equivalently, in the AC frequency $\Omega$. 

As a result of this expansion for the dot occupation one can show that the power developed by the AC source has a purely AC (Born-Oppenheimer)
component $P_{\rm cons}(t)$ associated to the reversible heat produced by the conservative forces, and a dissipative 
component $P_{\rm diss}(t)$ with a non-zero time average.  
The last term of Eq.~ (\ref{nd}) is associated to the frictional (dissipative) component of the force.
 In fact, by substituting Eq. (\ref{nd}) into Eq. (\ref{pac}) we find  
 \begin{equation}
 P(t)=P_{\rm cons}(t) +P_{\rm diss}(t)\,,
 \end{equation}
 with
\begin{eqnarray}\label{pacad}
P_{\rm cons}(t) &=& e \sum_{\sigma} n^f_{d \sigma} (t) \dot{V}_g(t), \nonumber \\
P_{\rm diss}(t) &= & e^2 \sum_{\sigma} \Lambda_{\sigma}(t)  [\dot{V}_g(t)]^2.
\end{eqnarray}
When performing the averages over one period $\tau= 2 \pi/\Omega$ 
\begin{equation}
\overline{P}_{\rm cons, diss}= (1/\tau)  \int_0^{\tau} dt P_{\rm  cons, diss}(t)\,,
\end{equation}
 for these two contributions to the 
power we can verify that 
$\overline{P}_{\rm cons}=0$ and $\overline{P}_{\rm diss} \geq 0$, as expected. 

We will analyze the adiabatic dynamics of the charge and energy at quantum dot and also the adiabatic regime of the charge and energy currents flowing in the normal leads. The latter can be carried out by recourse to 
non-equilibrium 
Green's function approach, as explained below. 

\subsection{Green's function approach}
We present the general expressions to calculate the relevant time-dependent mean values of the observables defined in the previous sections  by using the nonequilibrium Keldysh-Floquet Green's function formalism following Refs. \onlinecite{Arrachea-floquet,dyson} but now generalizing to the Nambu basis. 

One of the observables we are interested in is the occupation of the quantum dot. In order to evaluate it, the  starting point is the definition of the occupation matrix, with elements 
\begin{equation}\label{ocmat}
n^{ij}_{d\sigma}(t) = -i \left[G^{<}_{d,\sigma}(t,t)\right]_{ij},
\end{equation}
 which is 
defined from those of the lesser Nambu-Keldysh lesser Green's function matrix 
\begin{eqnarray}
\hat{G}^<_{d,\sigma}(t,t')=i
\begin{pmatrix}
\langle d^\dagger_{\sigma}(t') d_{\sigma}(t)\rangle &  \langle d_{\overline{\sigma}}(t') d_{\sigma}(t)\rangle \\
 \langle  d^\dagger_{\sigma}(t') d^\dagger_{\overline{\sigma}}(t)\rangle &  \langle d_{\overline{\sigma}}(t') d^\dagger_{\overline{\sigma}}(t)\rangle \\
\end{pmatrix}.
\end{eqnarray}
Here the upper (lower) signs correspond to spins $\uparrow$ and $\downarrow$, respectively, while $\overline{\sigma}$ denotes spin orientation opposite to $\sigma$. 
Particularly important for our purposes are the matrix elements 
\begin{equation}\label{exnd}
n_{d\sigma}(t) = {n}_{d\sigma}^{11} , \;\;\;\;\;\;\; \eta_{d\sigma}(t)= {n}_{d\sigma}^{12},
\end{equation}
which define, respectively, the population of the dot with electrons with spin $\sigma$ and with pairs induced at the quantum dot by proximity to the superconducting lead.

The lesser Green function matrix $\hat{G}_{d,\sigma}^<(t,t^{\prime})$ satisfies the Dyson equation
\begin{equation}\label{gless}
\hat{G}_{d,\sigma}^<(t,t^{\prime})= \int d t_1  d t_2 \hat{G}_{d,\sigma}^r(t,t_1) \hat{\Sigma}^<(t_1-t_2) \hat{G}_{d,\sigma}^a(t_2,t^{\prime}),
\end{equation}
where $\hat{G}_{d,\sigma}^r(t,t_1)= [\hat{G}_{d,\sigma}^a(t_1,t)]^{\dagger}$ are the retarded and advanced Green functions of the dot while $\hat{\Sigma}^<(t_1,t_2)$ encodes the coupling self-energy for the dot-reservoir.
The Fourier transform for the coupling self-energy reads $\hat{\Sigma}^<(\varepsilon)= i f (\varepsilon) \hat{\Gamma}(\varepsilon)$ and $\hat{\Gamma}(\varepsilon)= - 2 \mbox{Im}\left[ \hat{\Sigma}_S(\varepsilon) + \hat{\Sigma}_N (\varepsilon) \right]$ which are the coupling self-energies for the normal and superconducting contact and $f(\varepsilon)=1/[1+\exp{\beta\varepsilon}]$  is the Fermi-Dirac  function with $\beta=1/k_BT$ being $T$  the temperature, and $k_B$ the Boltzman constant (we recall that  we have assumed $\mu_N=\mu_S=0$). 

Another observable we need is the charge  current at the normal lead, which can be expressed in terms of Green's functions as  follows
\begin{eqnarray}\label{chargecurrent}
I_N(t)=- \frac{2e}{\hbar}  \sum_{\sigma} \int dt_1 \int \frac{d \varepsilon}{ 2 \pi} e^{-i \varepsilon (t_1- t)/\hbar } \\ \times \nonumber
\mbox{Re} \left[ \hat{G}_{d,\sigma}(t,t_1) \hat{\Sigma}_N^<(\varepsilon) + \hat{G}_{d,\sigma}^<(t,t_1) \hat{\Sigma}^a_N(\varepsilon) \right]_{11}.
\end{eqnarray}
Similarly, the two terms  of Eq. (\ref{qal}) defining the heat flux into the $N$ reservoir $\dot{Q}_{N}(t)$ can also be expressed in terms of Green's functions
\begin{eqnarray}\label{enercurrents}
J^E_{N}  & = &  - \frac{2}{\hbar}  \sum_{\sigma} \int dt_1 \int \frac{d \varepsilon}{ 2 \pi} e^{-i \varepsilon (t_1- t)/\hbar } \;  \varepsilon \; \\ &&\times \nonumber 
\mbox{Re} \left[ \hat{G}_{d,\sigma}(t,t_1) \hat{\Sigma}_N^<(\varepsilon) + \hat{G}_{d,\sigma}^<(t,t_1) \hat{\Sigma}^a_N(\varepsilon) \right]_{11} , \nonumber \\ 
J^E_{cN}  &=&  \frac{2}{\hbar} \int \frac{d \varepsilon}{ 2 \pi} f(\varepsilon) \;  \mbox{Re}  \left[  \partial_{t} \hat{G}_{d,\sigma}(t,\varepsilon) \hat{\Gamma}_N (\varepsilon) \right]_{11}. 
\end{eqnarray}

Since the retarded and advanced dot Green functions depend on two times it is convenient to work in the mixed representation 
\begin{equation}
\hat{G}_{d,\sigma}^r(t,t_1)= \int \frac{d \varepsilon}{2 \pi} \hat{G}_{d,\sigma}^r(t,\varepsilon)e^{-i \varepsilon (t-t_1)/\hbar}\,
\end{equation}
where in terms of Fourier components reads
\begin{equation}\label{gret}
\hat{G}_{d,\sigma}^r(t,t_1) = \sum_n e^{-i n \Omega t} \int \frac{d \varepsilon}{2 \pi} \hat{G}_{d,\sigma}^r(n,\varepsilon)e^{-i \varepsilon (t-t_1)/\hbar}.
\end{equation}
Similarly, the AC electrical field reads as follows in the Fourier representation,
$\hat{V}(t)=\sum_{n \neq 0} \left[ \hat{V}^+_n e^{-i n \Omega  t}+\hat{V}^-_n e^{i n \Omega t} \right]$. Here $\hat{V}^{\pm}_n$ are matrices in Nambu space with non-vanishing matrix elements, respectively, 
\begin{equation}
\left[\hat{V}^+_n\right]_{11}=\frac{e}{\tau}  \int_0^{\tau} e^{i n \Omega t} V_g(t), \;\;\; \left[\hat{V}^-_n\right]_{22}=-\frac{e}{\tau}  \int_0^{\tau} e^{-i n \Omega t} V_g(-t).
\end{equation}
Finally, the Fourier-transform in $t-t_1$ of the Green's function obeys the Dyson equation
\begin{equation}\label{dyr}
\hat{G}_{d,\sigma}^r(t,\varepsilon) = \hat{G}_{0}(\varepsilon) + \sum_{s=\pm} \sum_n e^{-i s n \Omega t} \hat{G}_{d,\sigma}^r(t,\varepsilon + s n \hbar\Omega) \hat{V}^s_n  \hat{G}_{0}(\varepsilon).
\end{equation}

\subsubsection{Adiabatic expansion of the Green's function}
 For the adiabatic dynamics  we just need a solution accurate upto ${\cal O}(\Omega)$ for  Eq. (\ref{dyr}). Expanding the rhs of this equation in powers of $\Omega$ leads to
\begin{equation}
\hat{G}_{d,\sigma}^r(t,\varepsilon) [\hat{G}_{0}(\varepsilon)^{-1}- \hat{V}(t)]= \hat{1}+ i \hbar \partial_{\varepsilon} \hat{G}_{d,\sigma}^r(t,\varepsilon) \frac{d \hat{V}(t)}{dt}.
\end{equation}
The explicit solution to this equation reads
\begin{eqnarray} \label{adia}
\hat{G}_{d,\sigma}^r(t,\varepsilon) \sim  & & \hat{G}_{f,\sigma}^r(t,\varepsilon) + i e \hbar \partial_{\varepsilon} \hat{G}_{f,\sigma}^r(t,\varepsilon)   \hat{G}_{f,\sigma}^r(t,\varepsilon) \dot{V}_g (t),
\end{eqnarray}
where $\hat{G}_{f,\sigma}^r(t,\varepsilon) = [\hat{G}_{0}(\varepsilon)^{-1}- \hat{V}(t)]^{-1}$ is the frozen dot Green's function. 

\subsubsection{The frozen dot Green's function}
The frozen Green's function corresponds to the equilibrium problem defined by the Hamiltonian frozen at the time $t$. It can be directly calculated by the equibrium 
 Dyson equation
\begin{equation}\label{gfro}
\hat{G}_{f,\sigma}^r(t,\varepsilon) \left[ \varepsilon \hat{1} - \hat{V}(t) - \hat{\Sigma}_N- \hat{\Sigma}_S \right] = \hat{1}.
\end{equation}
We recall that $\hat{\Sigma}_N$ is the self-energy describing the coupling between the quantum dot and the normal reservoir and the matrix $\hat{\Sigma}_S$ describes the coupling to the superconducting one.
In analogy to Eqs. (\ref{exnd}), we define the frozen occupation matrix, with elements 
$[\hat{n}_{f,\sigma}(t)]_{ij}= -i [\hat{G}_{f,\sigma}^<(t,t)]_{ij}$, where the lesser Green's function matrix satisfies
\begin{equation}
\hat{G}_{f,\sigma}^<(t,\varepsilon)=\hat{G}_{f,\sigma}^r(t,\varepsilon) \hat{\Sigma}^<(\varepsilon) \left[\hat{G}_{f,\sigma}^r(t,\varepsilon) \right]^*
\end{equation}
In our calculations, we will use the following matrix elements, which define the frozen  occupation of the quantum dot by particles and  by induced pairs
\begin{equation}
n_{d\sigma}^f=\hat{n}_{f\sigma}^{11}(t),\;\;\;\;\;\;\;\;\; \eta_{d\sigma}^f(t)=\hat{n}_{f\sigma}^{12}(t).
\end{equation}

The simplest model for the reservoirs corresponds to a constant density of states for the single particle energies. This results in the following self energy for the normal lead
\begin{equation}
\hat{\Sigma}_N = \left( \begin{array}{cc} -i \Gamma_N/2 & 0  \\
0 & -i \Gamma_N/2  \end{array}   \right).
\end{equation}
Similarly, the self-energy for the superconducting lead reads
\begin{equation}
\hat{\Sigma}_S =- \frac{\Gamma_S \left\{\varepsilon \; \theta(\Delta - |\varepsilon|) + i \; |\varepsilon|  \; \theta(|\varepsilon|- \Delta) \right\} }{2 \sqrt{|\Delta^2 - (\varepsilon + i 0^+)^2|}} 
\left( \begin{array}{cc} 1 &  \Delta/\varepsilon \\
    \Delta/\varepsilon  & 1 \end{array}   \right).
\end{equation}
Within this model for the self-energy, it is easy to show that the Green's function satisfies the properties presented in Appendix (\ref{prop}).  

In order to get explicit expressions we follow Ref. [\onlinecite{arrsup}]. We name $\left[ \hat{G}_{f,\sigma}^r(t,\varepsilon) \right]_{11} =- \left[ \hat{G}_{f,\sigma}^r(t, - \varepsilon) \right]_{22}^* = 
G(t, \varepsilon)$ and  $\left[ \hat{G}_{f,\sigma}^r(t,\varepsilon) \right]_{12}=\left[ \hat{G}_{f,\sigma}^r(t,\varepsilon) \right]_{21}^*= F(t, \varepsilon)$. 
In this case we get
\begin{eqnarray}\label{sol}
G (t, \varepsilon) & = & \frac{1}{\varepsilon  -eV_g(t) -  \Sigma_{\rm eff}(t, \varepsilon) }
\end{eqnarray}
where we have defined an effective self-energy $\Sigma_{\rm eff}(t,  \varepsilon)  =   \Sigma^G(\varepsilon)+ \Sigma^F(\varepsilon)^2 \overline{g}(t, \varepsilon)$ with the help of 
\begin{equation}
\Sigma^{G}(\varepsilon)=\sum_{\alpha=N,S}\left[\hat{\Sigma}_{\alpha}(\varepsilon)\right]_{11},\quad\quad \Sigma^{F}(\varepsilon)=\sum_{\alpha=N,S}\left[\hat{\Sigma}_{\alpha}(\varepsilon)\right]_{12}
\end{equation}
 and $\overline{g}(t, \varepsilon)  = 1/[\varepsilon + eV_g(-t) + \Sigma^{G}(-\varepsilon)^* ]$. Finally, the anomalous propagator reads
\begin{equation}\label{f}
F(t, \varepsilon) = - G(t, \varepsilon) \Sigma^F(\varepsilon) \overline{g}(t, \varepsilon)\,,
\end{equation}

\section{Instantaneous Joule law for the dot dynamics}
Introducing the adiabatic expansion of the Green's function of Eq. (\ref{adia}) into the definition of the occupation of Eq. (\ref{ocmat}) we can identify the two contributions to the adiabatic dynamics of the occupation of the quantum dot. The frozen contribution is determined from the frozen Green's function. 
 Conveniently, we define
\begin{equation}\label{rho}
\hat{\rho}_{f, \sigma}(t,\varepsilon)  =i \left[\hat{G}_{f, \sigma} (t,\varepsilon) -  \hat{G}_{f,\sigma}(t,\varepsilon)^* \right]\,,   
\end{equation}
in terms of which the frozen occupation matrix reads
\begin{equation} \label{nf}
n^{ij}_{f \sigma}(t) = \int \frac{ d\varepsilon}{ 2 \pi} f(\varepsilon) 
\left[ \hat{\rho}_{f,\sigma}(t,\varepsilon) \right]_{ij}.
\end{equation}
The coefficient of the linear contribution in $\dot{V}_g$ of Eq. (\ref{nd}) becomes
\begin{eqnarray} \label{lambda}
& & \Lambda_{\sigma}(t)  = -2 \hbar \mbox{Im} \left[  \int \frac{d \varepsilon}{2 \pi}  f(\varepsilon) \partial_{\varepsilon} \hat{G}_{f,\sigma}(t,\varepsilon)   \hat{\rho}_{f, \sigma}(t,\varepsilon) \right]_{11} \nonumber \\
& = &  - \frac{ \hbar}{2} \int  \frac{d \varepsilon}{2 \pi} \partial_{\varepsilon}  f(\varepsilon) \left\{ \left[   \hat{\rho}_{f, \sigma}(t,\varepsilon) \right]_{11}^2 + \left[   \hat{\rho}_{f, \sigma}(t,\varepsilon) \right]_{12}^2  \right\}.
\end{eqnarray}
Notice that  in the last step we have integrated by parts and used $ \left[\hat{G}_f(t,\varepsilon) \right]_{12}= \left[ \hat{G}_f(t,\varepsilon) \right]_{21}$, which implies $\left[ \hat{\rho}_{f, \sigma}(t,\varepsilon) \right]_{12}= \left[ \hat{\rho}_{f, \sigma}(t,\varepsilon) \right]_{21}$. Hence, this coefficient can be split in two components as 
$\Lambda_{\sigma}(t)= \Lambda_{11,\sigma}(t)+\Lambda_{12,\sigma}(t)$ being at zero temperature 
\begin{equation}\label{Lambdaij}
\Lambda_{ij, \sigma}(t)= \frac{\hbar}{4 \pi }\left[   \hat{\rho}_{f, \sigma}(t,0) \right]_{ij}^2.
\end{equation}

Now we evaluate the dissipative power from Eq. (\ref{pacad}) by using Eq. (\ref{Lambdaij}). We see that this quantity also has two components, associated to those of
$\Lambda_{\sigma}(t)$.
We will show below that the component related to $\Lambda_{11,\sigma}(t)$ follows a normal  instantaneous Joule law and we name it $P^N_{\rm Joule}(t)$, while
the one related to $\Lambda_{12,\sigma}(t)$ is named $\tilde{P}_{\rm Joule}(t)$ and follows an {\em anomalous Joule law},
\begin{equation} \label{pjoule1}
P_{\rm diss}(t)=P^N_{\rm Joule}(t)+ \tilde{P}_{\rm Joule}(t),
\end{equation}
with
\begin{eqnarray}
P^N_{\rm Joule}(t)  & = &  \frac{e^2 \hbar }{4 \pi } \sum_{\sigma}    \left[\hat{\rho}_{f, \sigma}(t,0) \right]_{11}^2 \dot{V}^2_g(t)  , \nonumber \\
\tilde{P}_{\rm Joule}(t)  & = & \frac{e^2 \hbar }{4 \pi } \sum_{\sigma}  \left[   \hat{\rho}_{f, \sigma}(t,0) \right]_{12}^2  \dot{V}^2_g(t) \,. 
\end{eqnarray}
In order to  make the Joule law explicit,  we proceed to  relate the two components of the dissipative power Eq. (\ref{pjoule1}) to the dot charge dynamics. To this end,
we analyze   
the time evolution of the  dot charge  up to ${\cal O}(\dot{V}_g(t) )$. This leads to the purely
AC charge current which reads
\begin{equation}\label{ic}
\frac{d{q}_{d\sigma}(t)}{dt}=e\dot{n}^f_{d \sigma}(t) = C_{\sigma}(t) \dot{V}_g(t).
\end{equation}
Here we can identify the non-linear capacitance of each spin channel $C_{\sigma}(t)=  e \partial n^f_{d \sigma}(t)/\partial V_g$. In addition, the dynamics of the
charge and heat involves the dynamics of the induced pairs at the quantum dot by proximity to the superconductor. The latter is 
$\eta_{d \uparrow }^f(t)=\langle d^{\dagger}_{\uparrow} d^{\dagger}_{\downarrow} \rangle = \langle  d_{\downarrow} d_{\uparrow} \rangle = -\eta_{d\downarrow }^f(t)$. The corresponding charge fluctuation reads
\begin{equation}\label{ic0}
\frac{d {q}^{\eta}_{d \sigma} (t)}{dt}= e \dot{\eta}^f_{d \sigma }(t) =  \pm \partial \eta^f_{d \uparrow}(t)/\partial \dot{V}_g(t),
\end{equation}
where the upper (lower) sign corresponds to $\sigma= \uparrow, \downarrow$, respectively.
Importantly, we get two contributions with opposite sign in (\ref{ic0}), which reflects the fact that a pair fluctuation implies a simultaneous flux  of electrons and holes. 
As a consequence, the net induced current between dot and reservoirs vanishes, although the process leads to energy dissipation in the form of a Joule law for the electrons and for the holes. Notice that each of the contributions
$\partial{\eta}_{d \sigma }^f(t)/\partial V_g$ 
can be positive of negative, depending on the occupation of the quantum dot. However, as they have opposite sign, the net contribution cancels when they are added.
For this reason, we  find it  convenient to define the "anomalous capacitance" as 
$\tilde{C}(t)= e |\partial \eta^f_{d \uparrow}(t)/\partial V_g|$ and redefine the induced-pair charge fluctuations as
\begin{equation}\label{ic1}
\frac{d\tilde{q}_{d \sigma} (t)}{dt}= \pm \tilde{C}(t) \dot{V}_g(t),
\end{equation}
which satisfies $\sum_{\sigma} d\tilde{q}_{d \sigma} (t)/dt = \sum_{\sigma} d {q}^{\eta}_{d \sigma} (t)/dt =0$.

In order to compute the capacitances we evaluate the dynamics of the dot charge  at first order in $\dot{V}_g(t)$. Then, starting from 
\begin{eqnarray}\label{dotdynamics}
\dot{n}_{d \sigma}^{ij}(t) &= &\int \frac{d\varepsilon}{2 \pi} f(\varepsilon) \frac{ d \left[ \hat{\rho}_{f,\sigma}(t,\varepsilon) \right]_{ij}}{dt} \nonumber \\
 &= & e \!\! \int  \frac{d\varepsilon}{2 \pi} \partial_{\varepsilon}f \left[ \hat{\rho}_{f,\sigma}(t,\varepsilon) \right]_{ij}\!\! \dot{V}_g,
\end{eqnarray}
and comparing Eq. (\ref{dotdynamics}) with Eqs. (\ref{ic}), and (\ref{ic1}) in the zero temperature limit we find
\begin{equation} \label{cc}
C_{ \sigma}(t) =  \frac{e^2}{2 \pi} \left[  \hat{\rho}_{f,\sigma}(t,0) \right]_{11}, \;\;\;\;\;\tilde{C}(t)= \frac{e^2}{2 \pi} \left| \left[  \hat{\rho}_{f,\uparrow}(t,0) \right]_{12} \right|.
\end{equation}
The dot charge dynamics [see Eq. (\ref{ic})] and the time evolution for the pair-density charge [see Eq. (\ref{ic1}) ] together with Eq. (\ref{cc}) can now be related to the 
normal and anomalous Joule components of the
dissipative power  [see Eq.(\ref{pjoule1})]  according to
\begin{eqnarray} \nonumber
P^N_{\rm Joule}(t) & = & 
R_0 \sum_{\sigma}    \left[ \frac{d q_{d\sigma}(t)}{dt}\right]^2, \nonumber \\
\tilde{P}_{\rm Joule}(t) & = &  R_0 \sum_{\sigma}   \left[ \frac{d \tilde{q}_{d\sigma}(t)}{dt} \right]^2, \label{pjoule}
\end{eqnarray}
with a constant and universal quantum resistance $R_0= h/2e^2$. 
While in the first term of Eq. (\ref{pjoule}) the label $\sigma$ represents fluxes of charges with different spin components, in the second term it actually represents the two opposite  charges for the electrons and the holes.
We notice that the above dynamics can be described by the circuit sketched in Fig. \ref{fig1}(b), which corresponds to a generalization of the RC circuit of a driven quantum dot connected to
a normal reservoir. There are four different channels that run in parallel, each channel has its own capacitance. We will see that the normal capacitance $C_{\sigma}(t)$ has contributions associated to normal transport as well as to Andreev processes, while the anomalous capacitance accounts for the induced Cooper pair fluctuation. The latter process involve opposite currents of electrons and holes, which do not produce any net current. Each of these channels dissipate energy in the
form a  Joule law with the universal B\"uttiker resistance $R_0$. This result holds for arbitrary amplitude of the driving potential provided that the driving frequency  is low enough and the reservoirs have $T=0$.

\section{Instantaneous Joule law at the normal contact }
We  recall that we are considering the chemical potential within the superconducting gap. This regime is interesting because the heat flux to the superconducting reservoir vanishes, which means that the dissipated energy flows only into the normal lead. In this situation 
we can get analytic expressions for the currents into the normal lead in the adiabatic regime. Our aim now is to verify that such heat flux also obeys
an instantaneous Joule law with B\"uttiker universal resistance $R_0$.  
We follow Refs. \onlinecite{Arrachea-floquet,dyson} to derive the charge and heat flow at the normal contact in the adiabatic approximation. Details are presented in 
 Appendix \ref{heatn}. We arrive at the expression for the heat current up to second order in  $\dot{V}_g(t)$ (equivalent to  up to ${\cal O}(\Omega^2)$. Such flux comprises two different contributions
 \begin{equation}
\dot{Q}_N(t)=\Lambda^{(1)}_{N}(t)(t) \dot{V}_g(t)+ \Lambda^{(2)} _{N}(t)\dot{V}_g(t)^2+\mathcal{O}[V_g(t)^3] +\cdot\,.
 \end{equation}
$\Lambda^{(1)}_{N}(t)$ is first order in the AC frequency $\Omega$ and 
vanishes at zero temperature. The other term is the second order contribution and reads for the zero temperature limit
  \begin{equation}\label{QN}
\dot{Q}_{N}(t)  \simeq 
 \frac{e^2 \hbar}{4 \pi}  \sum_{\sigma}  \{ \left[ \rho_{f,\sigma}(t,0)\right]_{11}^2 + \left[ \rho_{f,\sigma}(t,0)\right]_{12}^2 \} \dot{V}_g(t)^2.
\end{equation}
Notice that Eq. (\ref{QN}) is, precisely, the dissipated power $P_{\rm Joule}(t)$ given by Eq. (\ref{pjoule1}).  This result implies that 
the dissipative power coincides with the heat flow expression in the normal contact. Besides, it is important to emphasize that such heat current at the normal lead has been computed considering the contribution of the energy reactance,  see second term of Eq. (\ref{qal}).  

Finally, we calculate the expression for the charge current at the normal electrode at zero temperature which is calculated from Eq. (\ref{ial}) and it reads
\begin{equation}\label{IN}
I_N(t)= \frac{e^2}{2 \pi} \sum_{\sigma} \left\{ \left[ \rho_{f,\sigma}(t,0)\right]_{11} + \left[ \rho_{f,\sigma}(t,0)\right]_{12} \right\}\dot{V}_{g}(t).
\end{equation}
This again confirms the instantaneous Joule law [cf. Eq. (\ref{QN}) and Eq. (\ref{IN})].

 Therefore, the analysis of the fluxes in the normal lead
 confirms the description of the dissipation in our setup in terms of a circuit composed by two parallel subcircuits,  each of them corresponds to a RC circuit composed by the usual capacitance $C$ and $R_0$ and the anomalous capacitance $\tilde{C}$ and again $R_0$, respectively. The circuit picture reflects the fact that the normal reservoir effectively receives the charge flowing through all the resistive elements depicted in Fig. \ref{fig1}(b) that comes from (i) the normal transmission,  (ii) the Andreev processes  and  finally (iii) the Cooper pair fluctuation. All these transport events are the result of quasiparticle excitations that lead to energy dissipation. In the next section we will analyze these contributions in more detail.

\section{Analysis of the capacitances and the dissipated power} 
We now show results illustrating  the behavior of the capacitances, which determine the behavior of  the charge and heat currents between the quantum dot and the reservoir.
Substituting the dot Green function [see Eq. (\ref{sol})] in the expression for the dot density of states [see Eq. (\ref{rho})] we explicitly see that the capacitance for each spin channel given by Eq. (\ref{cc}) has two different contributions at zero temperature:
\begin{equation}
C_{\sigma}(t) \equiv C(t)= C_{\rm N}(t)  + C_{\rm And}(t)\,,
\end{equation}
We identify them to normal ($C_{\rm N}$) and Andreev-type processes ($C_{\rm And}$). 
They can be expressed in terms of the Green functions and self-energies previously defined [see Eq. (\ref{sol})], as follows
\begin{eqnarray}\label{cap}
C_{\rm N}(t) & = & \frac{e^2}{2 \pi} \Gamma^G(0)  |G(t,0)|^2 , \nonumber \\
C_{\rm And}(t) &= & \frac{e^2}{2 \pi} \Gamma^G(0) |G(t, 0) \Sigma^F(0)  \overline{g}(t, 0)|^2,
\end{eqnarray}
with $\Gamma^G(0)= -2 \mbox{Im}[\Sigma^G(0)]$.  Notice that the normal contribution is directly related to the normal part of the spectral function and exactly reduces to the capacitance of the quantum dot
connected to a single normal lead in the limit of vanishing coupling to the superconducting one. Instead the contribution $C_{\rm And}(t)$ is proportional to the coupling to superconducting the lead and involves  
high order scattering processes, characteristic of the Andreev reflection. The anomalous capacitance is 
\begin{equation}
\tilde{C}(t)= \frac{e^2}{\pi}| \mbox{Im}[F(t,0)] |.
\end{equation}
The latter is proportional to the absolute value of the spectral function of the anomalous Green's function [see Eq. (\ref{f})], which is positive (negative) for $V_g(t) >0$ ($V_g(t)<0$).

\begin{figure}
\includegraphics[width=\columnwidth]{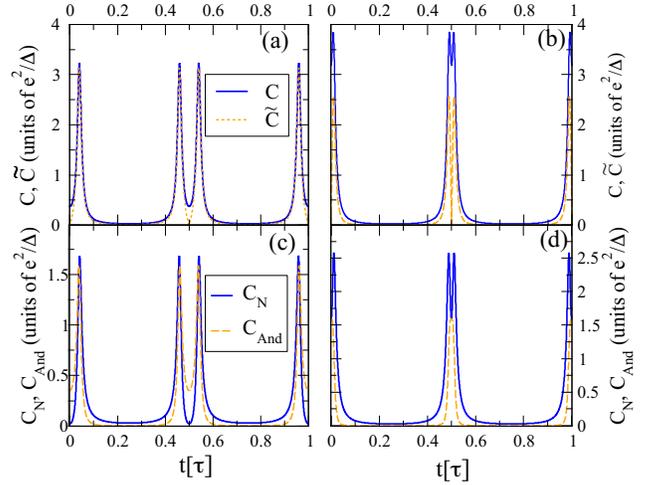}
\caption{\label{fig2} Capacitances of the driven quantum dot for $\Gamma_N= \Gamma_S/4=\Delta/10$ (left panels) and $\Gamma_N= \Gamma_S=\Delta/10$ (right panels). Times are expressed in units of the driving period. The amplitude of the driving gate voltage is $V_0=0.8 \Delta$. Upper panels: Solid lines and dots correspond to $C(t)$ and $\tilde{C}(t)$. Lower panels: Solid lines and dots correspond to $C_{\rm N}(t)$ and
$C_{\rm An}(t)$.} 
\end{figure}

The behavior of the different capacitances is illustrated in Fig. \ref{fig2}. Two different cases are shown, namely, the superconducting dominant case when $\Gamma_S>\Gamma_N$ (left panel in Fig. \ref{fig2} with $\Gamma_N=\Gamma_S/4$) and when both tunnel couplings are equal $\Gamma_N=\Gamma_S$ (right panel in Fig. \ref{fig2}). In the simplest situation where
the quantum dot is coupled only to the normal electrode and without driving, there is a single
level at the Fermi energy  $\mu_N=0$.  The additional coupling to the superconducting electrode  induces local pairing correlations in the quantum dot. Then, the original single dot level splits into two Andreev quasiparticle states in which the magnitude of the splitting depends on the relative value of $\Gamma_S/\Gamma_N$. Since the behavior of the capacitance is determined by the spectral properties of the quantum dot, these features are clearly identified  in Fig. \ref{fig2}.  In fact, for the superconducting dominant case the dot spectral density exhibits a larger level splitting in comparison to the case where both lead-dot couplings are similar. As a function of time, the gate voltage moves upwards and downwards. The Andreev quasiparticle energy levels and the capacitances have weights when the dot spectral functions have weight
at the Fermi energy $\mu=0$. Besides, we observe that the normal capacitance
 $C_{\rm N}(t)$ follows the profile of the Andreev levels, while the capacitance associated to Andreev reflection processes shows an additional weight between the two Andreev peaks. 
 We observe that the anomalous capacitance follows the spectral features of the anomalous Green function with resonances at the Andreev quasiparticle states. Besides, the anomalous Green function changes sign every time that $V_g(t)=0$, hence $\tilde{C}(t)=0$ at those times.  
 
\begin{figure}
\includegraphics[width=\columnwidth]{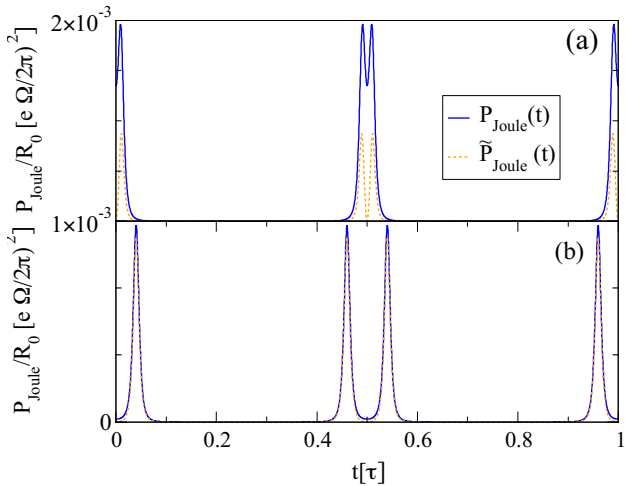}
\caption{\label{fig3} Dissipated power of the driven quantum dot
 for  $\Gamma_N= \Gamma_S=\Delta/10$ (top) and $\Gamma_N= \Gamma_S/4=\Delta/10$ (bottom). Solid lines and dots correspond to the contribution of the ordinary, $P_{\rm Joule}^N(t)$ and anomalous 
 $\tilde{P}_{\rm Joule}(t)$ components of the Joule law. Other details are the same as for Fig. \ref{fig2}.} 
\end{figure}

 Every time $C(t)$ and $\dot{V}_g(t)$ are finite, a charge current establishes between the quantum dot and the normal lead. This current has normal and Andreev components for each spin component leading to a net flux 
\begin{equation}
\sum_{\sigma} \frac{d{q}_{\sigma}(t) }{dt} = 2 \left[ C_{\rm N}(t) + C_{\rm And}(t) \right] \dot{V}_g(t).
\end{equation}
This flux leads to dissipation of energy following the Joule law 
\begin{equation}
P_{\rm Joule}^N(t) = 2 R_0 \left[ C_{\rm N}(t) + C_{\rm And}(t) \right]^2 \dot{V}_g(t)^2.
\end{equation}
The contribution due to the fluctuation of the induced pairing leads  to opposite particle and hole fluxes described by Eq. (\ref{ic1}) and has an associated  net vanishing current, 
\begin{equation}
\sum_{\sigma} \frac{d\tilde{q}_{d\sigma}(t)}{dt} =  0.
\end{equation} 
This process contributes, however, to the dissipation of energy, in the form of an anomalous Joule law 
\begin{equation}
\tilde{P}_{\rm Joule}(t) = 2 R_0 \tilde{C}(t)^2 \dot{V}_g(t)^2.
\end{equation}
  The different contributions to the dissipated power are shown in Fig. \ref{fig3}. Both contributions are peaked at the times where the energy levels of the Andreev states get aligned with the chemical potential of the leads.
  Due to the contribution of the Andreev capacitance, there is a finite current and Joule dissipation in the time intervals between these peaks in $P_{\rm Joule}(t)$. The anomalous dissipation $\tilde{P}_{\rm Joule}(t)$
  due to the disruption or formation of induced pairs vanishes exactly at the center of the gap between the pair of Andreev peaks.

\section{Conclusions} 
We have investigated the charge and energy dynamics of a driven quantum dot in contact to superconducting and normal leads. We have focussed on the adiabatic regime, relevant for
low frequency driving, with reservoirs at $T=0$. We have derived the dissipative power from (i) the dot charge dynamics and equivalently from (ii) the heat flow at the N contact that accounts the reactance contribution from the tunneling barriers. Besides, the charge current is calculated from (i) the time derivative of the dot charge and from (ii) the charge current flow exiting the normal contact. For both cases a dynamical Joule law is established leading to an universal nonlinear charge resistance $R_0=h/2e^2$.  In this scenario we have shown that the Joule dynamics law may be
described in terms of the RC circuit of Fig. \ref{fig1}(b). According to Fig. \ref{fig1}(b) the capacitance $C$ takes into account the normal and Andreev processes whereas  $\tilde{C}$ accounts for the generation of pairs. Remarkably, the current due to the pair generation vanishes as a result from the cancellation of electron and hole flows. However, it is important to highlight that such pair fluctuation processes contribute to the heat dissipation through an instantaneous Joule heating. 

\section{Acknowlegments} 
LA thanks Alfredo Levy Yeyati for reading the manuscript and interesting comments. LA thanks the support of the Alexander von Humboldt foundation, as well as CONICET, UBACyT 
and MinCyT from Argentina. 

\appendix

\begin{widetext}
\section{Properties and identities of the frozen Green's function}\label{prop}
The frozen Green's function satisfies the following properties
\begin{eqnarray}
 \hat{\rho}_{f, \sigma}(t,\varepsilon) & = &i \left[\hat{G}_{f,\sigma}(t,\varepsilon) -  \hat{G}_{f,\sigma}(t,\varepsilon)^* \right]  
 =    \hat{G}_{f,\sigma}(t,\varepsilon) \hat{\Gamma}(\varepsilon)  \hat{G}_{f,\sigma}(t,\varepsilon)^\dagger, \\
 \partial_{\varepsilon} \hat{G}_{f,\sigma}(t,\varepsilon) & \simeq  & - \hat{G}_{f,\sigma}(t,\varepsilon)^2  \nonumber \\
\frac{d  \hat{G}_{f,\sigma}(t,\varepsilon)}{dt}  & = & \hat{G}_{f,\sigma}(t,\varepsilon)  \hat{G}_{f,\sigma}(t,\varepsilon) e \dot{V}_g(t)   \simeq  -e \partial_{\varepsilon} \hat{G}_f(t,\varepsilon)  \dot{V}_g(t)
\end{eqnarray}
where $\hat{\Gamma}(\varepsilon)= i \left[\hat{\Sigma}^r(\varepsilon) - \hat{\Sigma}^r(\varepsilon)^* \right]$. In the last identities, we have assumed that we can neglect the dependence on $\varepsilon$ of $\hat{\Sigma}$, which is a valid assumption for models of reservoirs introduced in Section III.D.2.

\section{Adiabatic expansion for the charge and heat currents into the $N$ reservoir for subgap driving}\label{heatn}
The charge and energy currents in the the normal lead are defined, respectively, in Eqs. (\ref{chargecurrent}) and  (\ref{enercurrents}).
Substituting Eq. (\ref{gret}) in these expressions and using identities for the Green functions along the same steps presented in Refs. \onlinecite{re1,re2} but expressed in the Nambu representation we get
\begin{eqnarray}\label{curn}
I_{N}(t) & = &\frac{e}{\hbar} \sum_{\sigma} \sum_l e^{- i l \Omega t} \int \frac{d \varepsilon}{2 \pi} 
\left\{ \hat{\Gamma}_{\rm N}(\varepsilon) \left[ i \hat{G}^*_{d,\sigma}(-l,\varepsilon) \left[ f(\varepsilon) - f(\varepsilon - l \hbar \Omega) \right] \right. \right. \nonumber \\
& &\left. \left. + \sum_n \sum_{\beta= N, S}  \left[ f(\varepsilon + n \hbar \Omega) - f(\varepsilon) \right] \hat{G}_{d \sigma} (l+n,\varepsilon) \hat{\Gamma}_{\beta}(\varepsilon)
\hat{G}^*_{d, \sigma} (n,\varepsilon) \right] \right\}_{11},\nonumber \\
\dot{Q}_{N}(t) &=& \frac{1}{\hbar}  \sum_{\sigma} \sum_l e^{- i l \Omega t} \int \frac{d \varepsilon}{2 \pi} 
\left\{ \hat{\Gamma}_{\rm N}(\varepsilon) \left[ i \hat{G}^*_{d,\sigma}(-l,\varepsilon) \left( \varepsilon - \frac{l  \hbar \Omega}{2} \right) \left[ f(\varepsilon) - f(\varepsilon - l \hbar \Omega) \right] \right. \right. \nonumber \\
& &\left. \left. -  \sum_n \sum_{\beta= N, S} \left( \varepsilon + (\frac{l}{2} + n) \hbar \Omega \right) \left[ f(\varepsilon + n \hbar \Omega) - f(\varepsilon) \right] \hat{G}_{d ,\sigma} (l+n,\varepsilon) \hat{\Gamma}_{\beta}(\varepsilon)
\hat{G}^*_{d, \sigma} (n,\varepsilon) \right] \right\}_{11}, 
\end{eqnarray}

In order to get the adiabatic  expansion for the currents, we have to introduce Eq. (\ref{curn}) the adiabatic expansion for the Green's function defined in Eq. (\ref{adia}) and the corresponding expansion for the Fermi-Dirac distribution function
\begin{equation}
f(\varepsilon - l \hbar \Omega)= f(\varepsilon) - l \hbar \Omega \frac{ \partial f}{\partial \varepsilon} + \frac{1}{2} (l \hbar \Omega)^2 \frac{ \partial^2 f}{\partial \varepsilon^2}.
\end{equation}
Then, we keep the terms of the  charge current upto 
${\cal}(\hbar \Omega)$
and of the
heat current in the first and second order in $ \hbar \Omega$. Here, we also use the fact that within the gap, the density of states of the superconducting lead vanishes, hence, 
${\Gamma}_S \sim 0$ for $|\varepsilon|<\Delta$.
The results are the following
\begin{eqnarray}
I_N^{(1)}(t) & = & e^2   \sum_{\sigma} \int \frac{d \varepsilon}{2 \pi} \partial_{\varepsilon} f(\varepsilon) \left\{\left[ \rho_{f,\sigma} (t,\varepsilon) \right]_{11}+ \left[\rho_{f,\sigma}(t,\varepsilon) \right]_{12} \right\}\dot{V}_g(t), \nonumber \\
\dot{Q}^{(1)}_{N}(t) & \simeq &- e \hbar \sum_{\sigma} \int \frac{ d\varepsilon}{2 \pi}  \varepsilon \; 
\partial_{\varepsilon} f(\varepsilon) \left\{ \left[ \rho_{f,\sigma} (t,\varepsilon) \right]_{11}+ \left[\rho_{f,\sigma}(t,\varepsilon) \right]_{12} \right\}\dot{V}_g(t),  \\
\dot{Q}^{(2)}_{N}(t) & \simeq &- \frac{\hbar}{2} \sum_{\sigma, \beta} \int \frac{ d\varepsilon}{2 \pi}  \partial_{\varepsilon} f(\varepsilon) 
 \left[ \partial_t  \hat{G}_{f,\sigma}(t,\varepsilon) \hat{\Gamma}_{\beta}(\varepsilon) \partial_t  \hat{G}^*_{f,\sigma}(t,\varepsilon) \right]_{11}  = 
- \frac{e^2 \hbar}{2} \sum_{\sigma}  \int \frac{ d\varepsilon}{2 \pi}  \partial_{\varepsilon} f(\varepsilon) \left\{ \left[ \rho_{f,\sigma}(t,\varepsilon)\right]_{11}^2 + \left[ \rho_{f,\sigma}(t,\varepsilon)\right]_{12}^2 \right\} \dot{V}_g(t)^2. \nonumber
\end{eqnarray}
In the last line, we have   dropped those contributions to $\dot{Q}^{(2)}_{\rm N}(t)$ that vanish at temperature $T=0$. 
Notice that $\dot{Q}^{(1)}_{\rm N}(t) $ also  vanishes at $T=0$.

\end{widetext}



\end{document}